# Phase-resolved Modeling Of Surf Zone Wave Transformation Over Idealized Rough Bottoms

Emile Guélard Ancilotti [1], Damien Sous [1], Denis Morichon [1], Patrick Marsaleix [2], Héloise Michaud [3] and Matthias Delpey [4]

[1] Université de Pau et des Pays de l'Adour, E2S UPPA, SIAME, Anglet, France.
[2] CNRS, LEGOS, Toulouse, France.
[3] Shom, Toulouse, France.
[4] SUEZ, Rivages Pro Tech, Bidart, France.
`emile.guelard-ancilotti@univ-pau.fr`

**Abstract.** Rough rocky seabeds dominate world's coastlines, making accurate modeling of wave transformation in such environments essential for understanding coastal processes and hazards. Wave breaking and friction are critical drivers of wave dissipation over rough seabeds, especially in the surf zone. Phase-resolved wave models, often relying on classical bed shear stress or canopy drag approaches, can fail to capture these processes due to limitations in their physical assumptions, especially for large roughness elements typical of rocky seabeds. The present study adapts the Bulk Canopy Drag (BCD) parameterization, inherited from vegetation and porous media studies, into the 3D non-hydrostatic phase-resolved SYMPHONIE model to simulate wave dissipation over rough seabeds. The model was tested against laboratory experiments (LEGOLAS), using controlled irregular wave forcing propagating over a rough ramp. Results reveal that turbulent drag dominates the outer surf zone, while inertial drag plays a key role in the inner surf zone. A combined optimization of these influences achieved strong agreement with experimental measurements under one irregular wave forcing and a specific bottom configuration. These findings underscore the importance of accurately incorporating both turbulent and inertial contributions into wave dissipation models for the rocky surf zone. The proposed BCD approach provides a promising framework for improving phase-resolved modeling of wave dynamics, with potential applications for more complex field conditions. Future work will aim to extend the approach to diverse roughness configurations and refine empirical coefficients for broader scalability and in situ applicability.

## 1    Introduction

Rough bottoms account for almost 70% of the world's coastlines ([1]). Understanding wave transformation and predicting wave-driven processes and hazards in rocky environments requires accurately representing wave transformation in the surf zone for rough bottoms. Wave breaking and friction are key processes to quantify wave dissipation in models for these types of environments, especially in the surf zone. The focus is placed here on 3D free surface phase-resolved models, which offer promising prospects to describe interaction between natural and artificial structures at the wave scale. Up to date, the description of roughness-induced dissipation in phase-resolved wave models remains mostly based on the classical bed shear stress, often requiring the use of empirical or site-specific bottom drag coefficients. The classical bed shear stress approach, which relies on the assumption of a thin boundary layer, is questionable in rocky environments where large roughness elements can produce significant drag form, especially in the surf zone ([2]). An alternative is to describe bottom drag using a canopy-type volume drag, widely applied in vegetation or porous media studies (e.g., [3]). Comparisons between bed stress and canopy drag approaches on very rough bottoms remain very sparse ([4]). We will focus here on the canopy approach which has shown promising results on coral reef applications, offering a comprehensive closure of wave and momentum dynamics ([5]).

The present study aims to develop a parameterization of roughness-induced wave dissipation into 3D non-hydrostatic phase-resolved wave models. In particular, we aim to identify the respective roles played by turbulent and inertial drag components throughout the surfzone. The selected strategy was to adapt the bulk canopy drag (BCD) approach inherited from vegetation and porous media studies (e.g., [3]) to rough seabed and discretized water column. The phase-resolved 3D model SYMPHONIE ([6]) was tested against a series of recent laboratory experiments on surf zone waves dissipation over a ramp and in the presence of idealized roughness, including very large roughness elements ([4]).

## 2 Materials and methods

### 2.1 BCD approach

Initially designed to predict large-scale ocean circulation, the 3D non-hydrostatic version of SYMPHONIE ([6]) is adapted to perform phase-resolved simulations of surface waves. This model provides a detailed and dynamic representation of the temporal and spatial evolution of the free surface. By contrast to depth-averaged models, the vertical discretization of the water column also allows accounting for the vertical structure of the flow, particularly useful to describe interaction with complex seabeds.

Wave dissipation due to seabed roughness is modeled using the Bulk Canopy Drag (BCD) approach, which introduces an additional drag force into the governing momentum equations. This force captures the effects of solid obstacles in the water column and is inherited from studies on the effects of porous media for non-stationary flow ([7]). Its complete expression, expressed as the extended Darcy-Forchheimer equation, is:

$$F_{BCD} = \rho \left( C_D |u| u + C_L u + C_M \frac{\partial u}{\partial t} \right) \tag{1}$$

where $C_D$ (in m$^{-1}$), $C_L$ (in s$^{-1}$) and $C_M$ (adimensional) are coefficients representing turbulent, laminar, and inertial influences, respectively. Empirical determination of these coefficients can be challenging, especially for optimization against observations. The Reynolds (Re) and Keulegan-Carpenter (KC) numbers are commonly used to identify the hydrodynamic regimes within the canopy and the dominant components controlling the drag in Eq. 1. These dimensionless numbers are defined as:

$$Re = \frac{u_c l_c}{\nu} \quad KC = \frac{u_c T}{l_c} \tag{2}$$

where $u_c$ is a characteristic velocity, $l_c$ a characteristic length, $\nu$ the kinematic viscosity, and $T$ the wave period. In the present study, $T$ is defined as the peak period $T_p$, $l_c$ as the roughness influence height, equal to four times the standard deviation of the bottom elevation ([8]), and $u_c$ as the "significant" orbital wave current at the half of this height inside the canopy, equal to the square root of the squared current module multiplied by two. Although blocs width or height can be directly used as the characteristic length in the present study, the influence height was chosen following [9] because it is based on a statistical metric of the topography variability that remains accessible for real rough seabed, in contrast to the geometrical dimensions of an idealized seabed configuration. The influence height is also used to determine the portion of the water column over which the BCD force is applied.

In the presence of large roughness (such as large gravels or rocks) and for the range of Re and KC numbers calculated from the simulation (Figure 1), [7] recommends focusing on turbulent and inertial influences. Thus, only those two influences were considered.

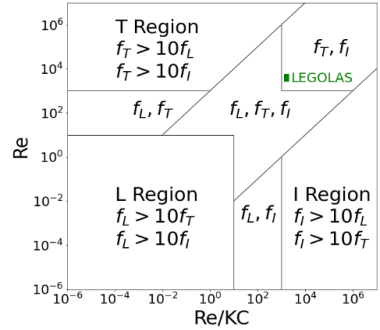

**Fig. 1.** LEGOLAS study case (green rectangle) in the framework of the Gu & Wang diagram (expressed as [7]) showing the relative influence of each expression of the drag ($f_L$ for the laminar one, $f_T$ for the turbulent and $f_I$ for the inertial) depending on Re and KC

## 2.2 Experimental setup

The model developments were compared to the LEGOLAS laboratory experiments ([9]) conducted in the CASH flume of SeaTech engineering school in Toulon (France). During these experiments, [9] measured variations in water surface elevation using 18 resistive probes, regularly spaced over a linear sloping bed (Figure 2). In the selected benchmark case, a JONSWAP irregular wave forcing with a $T_p$ of 1.2 s and a $H_s$ of 0.061 m was generated using a piston wave maker. Two seabed configurations were tested here: the smooth one, to validate the representation of wave transformation due to wave breaking only, and the rough one, based on a quincux configuration of 0.03 m cubes, with a roughness influence height of 0.044 m.

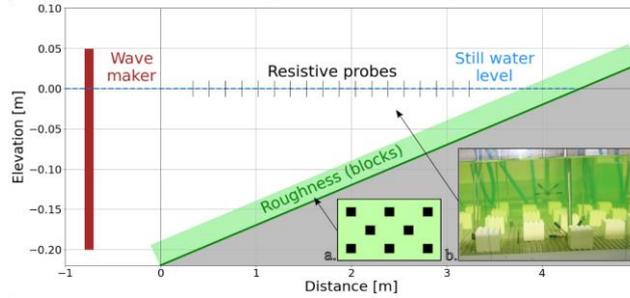

**Fig. 2.** Cross-shore scheme of the LEGOLAS experimental wave flume with (a.) the chosen blocs planar pattern on the ramp and (b.) a picture of the water column

### 2.3    Numerical setup and analysis

The numerical domain was defined as an 8 m long wave flume (extending by 2 m off-shore the experimental channel length to stabilize wave generation), 0.15 m wide and 0.22 m deep. The wave flume presented a linear slope of 1/20 starting 3 m from the wave generator. The horizontal resolution was 0.03 m and the time step was 0.01 s. Ten sigma layers, varying with the bathymetry and the free surface elevation variations, divided the water column. The geometric configuration was fully alongshore uniform, so that the analysis was carried out in a 2DV cross-shore framework. Lateral boundaries were defined as periodic and a flather condition was imposed at the wave maker open boundary.

Numerical sea surface elevation time series were extracted at each probe location and significant wave height ($Hs$) profiles were computed by integrating spectra at each location over the frequency band of gravity waves ($0.5f_p < f < 3f_p$, where $f_p$ is the peak frequency).

The first step was to calibrate the breaking parameters of the model without roughness-induced drag force against the smooth bottom configuration, to ensure the quality of the wave breaking representation. Once optimized, breaking parameters were kept constant for the rough case study, allowing us to isolate the roughness contribution to frictional dissipation.

Three model configurations for the rough seabed were compared:

- the sole turbulent component, with a constant optimized $C_D$ of 10 m⁻¹. Laminar and inertial drag components were neglected with coefficients values of 10⁻⁵,
- the sole inertial component, with a constant optimized $C_M$ of 0.5. Laminar and turbulent drag components were neglected with coefficients values of 10⁻⁵,
- the combination of turbulent and inertial influences, with constant optimized $C_D$ and $C_M$ of 10 m⁻¹ and 0.5. The laminar turbulent drag component was neglected imposing coefficient value of 10⁻⁵ s⁻¹.

Root mean square error (RMSE) and Willmott Index (WI) were computed to compare the results with the corresponding experimental measurements.

## 3 Results

The model results are compared with experimental cross-shore $Hs$ profiles, as shown in Figure 3. Four profiles illustrate the impact of different drag influences along the cross-shore profile:

- o The blue line represents the cross-shore profile of modelled $Hs$ compared to observations (blue crosses) for the smooth case. A good agreement is observed, with RMSE of 0.0024 m and WI of 0.988, demonstrating the performance of the model in reproducing the sole breaking-induced dissipation.
- o The green line shows the cross-shore profile of modelled $Hs$ for the case with turbulent influence solely. When compared to experimental measurements on the rough ramp (orange crosses), the model achieves an RMSE of 0.0041 m and a WI of 0.966.
- o The red line represents the cross-shore profile of modelled $Hs$ for the case with inertial influence solely, resulting in an RMSE of 0.0022 m and a WI of 0.992 compared to experimental measurements on the rough ramp.
- o Finally, the yellow curve corresponds to the $Hs$ profile obtained by combining turbulent and inertial influences. This configuration provides the best agreement with experimental measurements on the rough ramp, achieving an RMSE of 0.0017 m and a WI of 0.995.

These comparisons demonstrate the specific impacts of the drag components, which vary across the beach profile. Prior to breaking or around the breaking point, waves are large, inducing strong orbital velocities, which should explain the dominant effect of turbulent drag when compared to the inertial component. Toward the end of the ramp, in very shallow waters, the wave height decreases but the wave asymmetry is expected to be strong, which may explain why inertial forces dominate. Further analysis is required to better understand the driving processes. Overall, these results suggest that the combined optimization of turbulent and inertial influences is critical for accurately modeling wave transformation over rough bottom in the surf zone. Moreover, laminar drag did not need to be considered in that scenario, as the model showed good results compared to experimental measurements using only turbulent and inertial drag components. However, it would be important to take it into account for scenarios showing lower Re numbers.

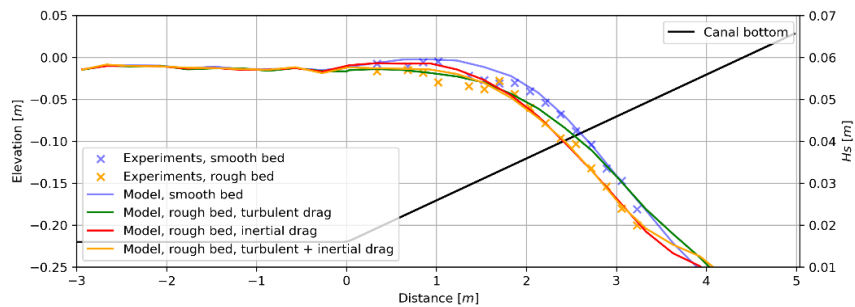

**Fig. 3.** Experimental and model $Hs$ profiles comparison (in colors) for the turbulent and/or the inertial contribution(s) to the drag force along the canal bottom (in black)

## 4    Conclusion and perspectives

This study highlights the potential of a 3D non-hydrostatic phase-resolved model, SYMPHONIE, to simulate wave transformation over rough seabeds in the surf zone. By incorporating the Bulk Canopy Drag (BCD) in the model, we parameterized roughness-induced wave dissipation, accounting for both turbulent and inertial drag influences. The comparison of numerical results with experimental data from the LEGOLAS laboratory experiments demonstrates the robustness of the proposed approach.

The results indicate that turbulent drag plays a dominant role in the outer surf zone, whereas inertial drag becomes more significant in the inner surf zone. This dual influence highlights the necessity of combining both contributions to accurately represent wave dissipation across varying depths in the surf zone. An optimized combination of these factors resulted in a good agreement with experimental measurements for a specific irregular wave forcing and bottom configuration.

These findings provide new insights for improving the representation of wave dissipation in rocky coastal environments, which are still poorly understood in current models. Future work will aim to extend the approach to more complex scenarios, including different roughness configurations and hydrodynamic conditions. Additionally, further investigation is needed to refine the empirical coefficients and explore the scalability of the approach for in situ applications. Although optimized drag coefficients were sufficient to represent turbulent and inertial influences for the chosen forcing and bottom configuration, further improvements could be achieved by expressing these coefficients according a local hydrodynamic expression depending on KC or Re ([4]) and a geometric influence depending on representative topographic metrics ([9]).